\documentclass[apj,iop,numberedappendix]{emulateapj}
\usepackage{apjfonts}
\usepackage{graphicx} 
\usepackage{amssymb}
\usepackage{amsmath}
\usepackage{color}

\newcommand{\be}{\begin{equation}}
\newcommand{\ee}{\end{equation}}
\newcommand{\bea}{\begin{eqnarray}}
\newcommand{\eea}{\end{eqnarray}}

\newcommand{\refeq}[1]{Eq.~(\ref{eq:#1})}

\newcommand{\reffig}[1]{Fig.~\ref{fig:#1}}          

\newcommand{\refapp}[1]{Appendix~\ref{app:#1}}
\newcommand{\vs}{\nonumber\\}       
\newcommand{\refsec}[1]{Section~\ref{sec:#1}}          

\renewcommand{\v}[1]{\vec{#1}}

\newcommand{\Aest}{\hat{A}}
\newcommand{\Best}{\hat{B}}

\newcommand{\tAest}{\tilde{A}}
\newcommand{\Mvir}{M_{\rm vir}}

\newcommand{\Msunh}{\,M_{\odot}/h}

\newcommand{\<}{\langle}
\renewcommand{\>}{\rangle}

\renewcommand{\k}{\kappa}
\renewcommand{\d}{\delta}
\newcommand{\D}{\Delta}

\renewcommand{\l}{\ell}
\newcommand{\vth}{\v{\theta}}

\newcommand{\s}{\sigma}

\newcommand{\vl}{\v{\l}}

\newcommand{\nbar}{\bar{n}}
\newcommand{\nobs}{n_{\rm obs}}

\renewcommand{\th}{\theta}
\newcommand{\g}{\gamma}

\newcommand{\A}{\mathcal{A}}

\newcommand{\nuth}{\nu_{\rm th}}
\newcommand{\Mmin}{M_{\rm min}}
\newcommand{\Np}{N_{\rm peak}}

\newcommand{\avg}[1]{\langle #1 \rangle}
\newcommand{\Var}{{\rm Var}}

\begin{document}

\shortauthors{Schmidt et al.}
\shorttitle{Cluster Finding with Weak Lensing}

\title{Weak Lensing Peak Finding: Estimators, Filters, and Biases}

\author{Fabian~Schmidt\altaffilmark{1} and Eduardo~Rozo\altaffilmark{2,3}}

\altaffiltext{1}{Theoretical Astrophysics, California Institute of
  Technology, M/C 350-17, Pasadena, California 91125, USA}
\altaffiltext{2}{Einstein Fellow, Department of Astronomy \& Astrophysics, The University of Chicago, Chicago, IL 60637, USA}
\altaffiltext{3}{Kavli Institute for Cosmological Physics, Chicago, IL
  60637, USA}

\begin{abstract}
Large catalogs of shear-selected peaks have recently become a reality.  
In order to properly interpret the abundance and properties of these
peaks, it is necessary to take into account the effects of the
clustering of source galaxies, among themselves and with the lens.  
In addition, the preferred selection of lensed galaxies in a flux- and 
size-limited sample leads to fluctuations
in the apparent source density which correlate with the lensing field
(\emph{lensing bias}).  
In this paper, we investigate these issues for two different choices of 
shear estimators which are commonly in use today:
globally-normalized and locally-normalized estimators.  
While in principle equivalent, in practice these estimators respond differently to 
systematic effects such as lensing bias and cluster member dilution.   
Furthermore, we find that which estimator is statistically superior depends on the
specific shape of the filter employed for peak finding; suboptimal choices
of the estimator+filter combination can result in a suppression of the
number of high peaks by orders of magnitude.  Lensing bias generally acts to
increase the signal-to-noise $\nu$ of shear peaks;  for high peaks the
boost can be as large as $\Delta \nu \approx 1-2$.  
Due to the steepness of the peak abundance function, these boosts can result
in a significant increase in the abundance of shear peaks. 
A companion paper \citep{paperA} investigates these same issues
within the context of stacked weak lensing mass estimates.
\end{abstract}


\maketitle
\section{Introduction}

The abundance of rare massive objects (clusters) in the Universe has emerged
as a powerful probe of cosmology \citep{Vikhlinin,RozoEtal10,MantzEtal09,HenryEtal09,vanderlindeetal10}.  
Many different techniques
can be used to find these clusters, such as optical identification,
X-rays, Sunyaev-Zeldovich effect, and gravitational lensing.  Among these, 
lensing stands out as being the least sensitive to the
complicated baryonic physics that govern galaxy formation
and the non-thermal processes that affect the dynamics of the intra-cluster
medium.  Consequently, the lensing signal is expected
to be the easiest to predict,
a fact that has fostered great interest in developing weak lensing 
cluster finders \citep{schneider96,HennawiSpergel,HamanaEtal2004}, and has even lead to the publication of
several lensing selected cluster samples \citep{MiyazakiEtal07,WittmanEtal06,GavazziSoucail}.
In practice, lensing selection suffers from significant projection effects:  
the lensing signal of a cluster can be enhanced by a favorable projection
of a triaxial halo; by associated mass distributions (substructure filaments); 
or by the chance superposition of large-scale structure 
along the line of sight.  Fortunately, such superpositions can be 
calibrated by relying on numerical
simulations, and weak lensing peak statistics remain a promising probe
for cosmological physics \citep{WangEtal,DietrichHartlap,KratEtal,marianetal10}.

Weak lensing shear measurements use the shapes of distant galaxies in order 
to statistically
extract the lensing signal.  To date, most work has assumed that
source galaxies are randomly distributed, whereas in practice we know
galaxies are clustered.  The apparent clustering receives
two contributions: one from intrinsic (physical) clustering of the
galaxies, henceforth referred to as \emph{source clustering}; the other from 
fluctuations in the galaxy density induced by lensing itself, via 
\emph{lensing bias} \citep[also known as magnification bias;][]{Broadhurst1994,paperI}.  
The intrinsic clustering of source galaxies acts to increase the noise in 
the shear measurement, while the lensing-induced fluctuations can bias 
the shear measurement if they are not properly accounted for.  In
fact, lensing bias can be seen as a probe of weak gravitational lensing
in its own right \citep[][Rozo and Schmidt, in preparation]{schneideretal00,vanwaerbekeetal10}.

In a companion paper \citep[][henceforth referred to as paper~I]{paperA}, 
we study similar issues for stacked
weak lensing analyses of groups and clusters.  The two papers are highly complementary,
as paper~I focuses on high signal-to-noise weak lensing mass calibration of 
objects that have been previously identified and then stacked, whereas in
this work we focus on identifying shear peaks in the low signal-to-noise 
regime.

Following paper~I, we discuss two possible filtered shear estimators, 
which differ in the way the estimator is normalized.  The first estimator uses a 
fixed normalization (e.g., \cite{MaturiEtal05,WangEtal}), and is therefore
sensitive to the overall modulation of the source density field by the
lensing signal.
The second estimator uses an individual normalization for each point in the 
sky (e.g., \cite{ MiyazakiEtal07}).  In this approach,
the fluctuations in the density of background galaxies are partially canceled
out, which reduces noise, but also shrinks the extra signal due to magnification.
Moreover, a location-based normalization estimator can lead to dilution
of the lensing signal if the source population is contaminated by cluster
galaxies (see paper~I for a more detailed discussion).

Throughout, we adopt a fiducial flat $\Lambda$CDM cosmology with
$h=0.7$, $\Omega_m=0.28$, $n_s=0.96$, and a power spectrum normalization
of $\s_8=0.85$ at $z=0$.  All masses are defined as $M_{200m}$, i.e. enclosing
an average density of 200 times the mean matter density.  
The source galaxies are assumed to follow the 
redshift distribution expected for the Dark Energy Survey
(DES)\footnote{http://www.darkenergysurvey.org/}, 
\be
\frac{dN}{dz} \propto z^2\exp\left[ -(z/z_0)^\beta \right]
\label{eq:dNdz}
\ee
with $z_0=0.5$ and $\beta=1.4$, and we assumed a density of $\nbar = 20$~arcmin$^{-2}$.  
This source density is slightly larger than that expected for DES, but smaller than
that of other future surveys such as Hyper Suprime-Cam or the Large Synoptic Survey Telescope (LSST).

\refsec{est} presents the shear estimators, and discusses 
how lensing bias impacts each of these estimators in turn.  We also
discuss the choice of filter function and optimal filter scale.  \refsec{res}  
presents the results on the statistics of lensing peaks.  
We conclude in \refsec{disc}.  Details on the lensing calculations, 
and two derivations regarding the variance of smoothed shear filters 
have been relegated to the appendix.

\section{Smoothed Shear Estimators}
\label{sec:est}

In this section, we focus on estimators of the form 
\be
\Aest(\vth) = \frac{1}{\bar n} \sum_i e_i W(\vth_i-\vth)\label{eq:Aest}.
\ee
where the sum is over all galaxies\footnote{Note that for simplicity, and in
order to keep results general, we have not included any galaxy weights, which
in practice will be used in particular if photometric redshifts are available.}, 
$\bar n$ is the mean source galaxy density, $e_i$ is the tangential component of the ellipticity of galaxy $i$ (with
respect to the relative position $\vth_i-\vth$), and $W$ is an arbitrary filter
which we assume is unity-normalized,
\be
\int d^2\th \:W(\vth) = 1.
\label{eq:Wnorm}
\ee
For the time being, we leave the filter unspecified.  The estimator \refeq{Aest}
has a \emph{fixed} normalization.  We consider the alternative choice of a
varying normalization in \refsec{Aprime}.  
 
In order to derive the statistical properties of $\Aest$, we proceed
as follows: first, we divide the sky into infinitesimal pixels of area $\D \Omega $
such that the number of galaxies in each pixel is either $0$ or $1$.  Letting
$n(\vth)$ be the galaxy density field on the sky, we can rewrite \refeq{Aest} as
\be
\Aest(\vth) = \frac{1}{\bar n} \sum_i n_i e_i W_i \D \Omega \label{eq:Aest1}.
\ee
where the sum is now over all pixels, and the index $i$ denotes that the quantity of interest
is evaluated at $\vth_i$.  For example, $n_i=n(\vth_i)$.  We now set
\be 
n_i = \bar n (1+\d_i)
\label{eq:n0}
\ee
where $\d_i$ is the galaxy density fluctuation (in this section, we assume
the fluctuations are purely Poisson).  Assuming $\avg{e_i}= g_i$ where
$g=\gamma/(1-\kappa)$ is the reduced (tangential) shear, we find that the expectation value of $\Aest$
is given by
\be
\avg{\Aest(\vth)} = \sum_i g_i W_i \D \Omega  = \int d^2\theta' g(\vth') W(|\vth-\vth'|).
\label{eq:Aestc}
\ee
For the second equality, we have let $\D\Omega\rightarrow 0$ (continuum
limit), and correspondingly replaced $\sum_i\D\Omega$ with $\int d^2\th$.  
Further, we have assumed that the source galaxy overdensity $\d$ is uncorrelated
with the shear.  In the following discussion, we will set $\vth=0$ without
loss of generality.  
We can compute the variance of $\Aest$ in a similar fashion.  In particular, using
\refeq{Aest1} and \refeq{n0} we find
\be
\Aest^2 = \sum_{ij} (1+\d_i\d_j)e_ie_jW_iW_j (\D\Omega)^2 
\label{eq:Aest2}
\ee
where have ignored terms proportional to $\delta$ since these will go to zero upon
taking the expectation value.  Neglecting any clustering of the source galaxies
for the moment, we have
\be
\avg{\d_i\d_j} = \frac{1}{\bar n \D\Omega} \d_{ij},
\ee
while the expectation value of the galaxy ellipticities takes the form (recall
that $e_i$ stands for the tangential component of the ellipticity)
\be
\avg{e_ie_j} = g_ig_j + \frac{\sigma_e^2}{2}\d_{ij}.
\ee
Here, $\sigma_e$ denotes the RMS (intrinsic) galaxy ellipticity of the sample.  
Using these expressions, taking the expectation value of
\refeq{Aest2}, and subtracting off $\avg{\Aest}^2$, we find
\be
\Var(\Aest) = \frac{1}{\bar n} \int d^2\theta\ W^2(\vth)\left( \frac{\sigma_e^2}{2}+g^2(\vth) \right).
\label{eq:Vsh0}
\ee
In many cases, $|g| \ll \sigma_e\sim 0.3$ and the term $g^2(\vth)$ in \refeq{Vsh0}
is often neglected.  In that case,
\be
\Var(\Aest) = \frac{\sigma_e^2}{2\bar n} \int d^2\theta\: W^2(\vth)
= \frac{\sigma_e^2}{2\bar n} \frac{1}{4\pi\Theta^2},
\ee
where for the second equality we have assumed a Gaussian shear filter of width 
$\Theta$ [\refeq{WGauss}], which is the standard result for the variance
of $\Aest$ \citep{vanWaerbeke2000}.

\subsection{Filter Functions}
\label{sec:filter}

A variety of smoothing kernels have been proposed for averaging the shear,
including top-hat, aperture mass (\cite{SchneiderEtal97}), and matched filters 
(\cite{MaturiEtal05,MarianBernstein}).  Some
differences in the filters are due to the various goals they were designed
to achieve; for example, to reduce contributions from small scales, or
large scales, or the mass-sheet degeneracy present in shear measurements. 

\begin{figure}[t!]
\centering
\includegraphics[width=0.48\textwidth]{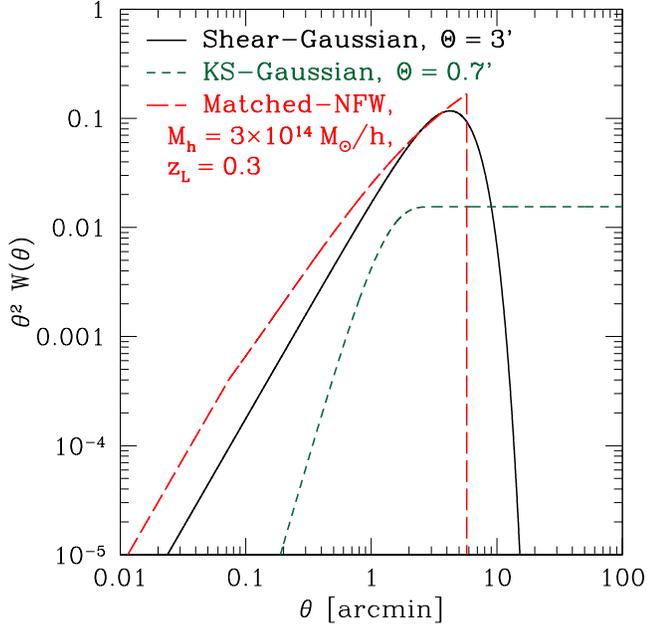}
\caption{Shear filter functions considered in this paper, as
a function of angular separation: 
Gaussian-smoothed shear filter;  Gaussian-smoothed convergence filter using the
method of \cite{KaiserSquires};  and matched NFW filter \citep{MarianBernstein}.
\label{fig:W}}
\end{figure}

In this paper, we will discuss three 
commonly used or proposed filters, which are shown in \reffig{W}.  It
is instructive to compare this figure with \reffig{kprof} in the appendix,
which shows the angular scales contributing to the lensing signal.  
Our first and perhaps simplest choice is a \emph{Gaussian shear filter},
\be
W(\th) = \frac{1}{2\pi\:\Theta^2} \exp\left( -\frac{\th^2}{2\Theta^2} \right ),
\label{eq:WGauss}
\ee
where $\Theta$ denotes the filter scale, which can be chosen to maximize
signal-to-noise for a given lens mass and redshift.  This filter is
also similar in shape to the filter presented in \cite{MaturiEtal05} designed to
reduce the contribution from unassociated large-scale structure.  

Another choice is a filter constructed using the method of \cite{KaiserSquires}
to yield an estimator of the Gaussian-smoothed \emph{convergence} $\kappa$
(``KS-Gaussian'').  This is equivalent to searching for peaks in smoothed
convergence maps \citep[e.g.,][but see caveats below]{HamanaEtal2004,WangEtal}.  The shear filter
function is given by:
\be
W(\th) = \frac{1}{\pi\theta^2} \left[ 1 - \left (1 + \frac{\th^2}{2\Theta^2}\right) 
\exp\left( -\frac{\th^2}{2\Theta^2} \right ) \right ],
\ee
where $\Theta$ is the width of the Gaussian convergence filter.  This
filter is usually used by pixelizing the sky in patches of a few
arcmin$^2$ area, and measuring the average shear in each pixel.  The resulting
shear map is then convolved with the KS-Gaussian filter.   While we do not 
explicitly consider pixelized estimators here,
under certain conditions the globally-normalized estimator \refeq{Aest} is 
equivalent to a pixelized estimator:  this is the case if either the
shear in each pixel is estimated with a global normalization (such as in
\refeq{Aest}), and pixels are weighted equally;  or if the shear in a pixel
is measured with a local normalization, and each pixel is then
inverse-variance weighted, i.e. weighted
by the number of galaxies in the pixel, in the further analysis.  

Note that
this filter is not normalizable through \refeq{Wnorm}.  Instead, the 
normalization is defined through the convergence filter.  
Due to the non-local relationship between shear and convergence, this filter
is quite different from the Gaussian shear filter.  As shown in \reffig{W},
small scales are downweighted, while in principle arbitrarily large scale 
scales are weighted equally ($\th^2\, W \sim$~const).  This is a consequence
of the mass-sheet degeneracy present in the shear, which can only be broken
by including very large scales.  We will see that due to these differences, 
a shear estimator using the KS-Gaussian filter behaves quite differently 
than an estimator using the other filters discussed here.

Finally, one can choose a filter matched to the expected signal of a
Navarro-Frenk-White (NFW, \cite{nfw96}) lensing halo.  
Following \cite{MarianBernstein}, we choose
\be
W(\th) = C\:g(\th),
\ee  
where $g$ is the reduced shear profile of an NFW halo (see \refapp{NFW}), 
and the constant 
is determined from the normalization constraint \refeq{Wnorm}.  
Note that as written, the filter is optimal for uniform noise and in the
absence of lensing bias;  it can easily be generalized to take into
account the effects discussed in this paper.  
In \cite{MarianBernstein}, the filter was truncated at the scale corresponding
to the virial radius (or $R_{200}$) of the halo (i.e. $W = 0$ for
$\th > \th_{\rm max}$);  in general, one
can vary the truncation scale of the filter.  \reffig{W} shows
that the matched-NFW filter (with truncation at $R_{200}$) is in fact
quite similar to the Gaussian shear filter.  

We will return to the choice of filter and optimal filter scale $\Theta$ in 
\refsec{filter2}.  

\subsection{Lensing bias and Source Clustering}
\label{sec:LB}

In the presence of foreground lensing matter, the number density of a flux- and size-limited
sample of background galaxies is affected by two effects: galaxies get pushed over the
flux and/or size threshold by lensing magnification, and their density
is diluted because the observed patch of sky is stretched (\cite{Broadhurst1994}).  
These two effects combine in such a way that the observed source galaxy
density field is related to the unlensed background density field via\footnote{In the weak lensing
literature, it is customary to linearize the magnification term $\mu^{q/2}\approx (1+q\kappa)$.  The parametrization \refeq{nobs1} is the general 
expression valid into the moderate lensing regime, see \cite{Broadhurst1994}.}
\be
\nobs(\vth)=\bar{n} [1+\d_g(\vth)]\mu(\vth)^{q/2}
\label{eq:nobs1}
\ee
where the magnification $\mu$ is given by\footnote{In the following, we will 
ignore the 
fact that the source redshift distribution $dN/dz$ itself
depends on $\theta$, and will always calculate $\mu$ assuming the
average $dN/dz$.  Since the lensing efficiency varies slowly with redshift, 
we expect such fluctuations to have negligible impact.}

\be\label{eq:mu}
\mu(\vth)=\frac{1}{(1-\k)^2-|\g|^2} = 1 + 2\k + 3\k^2 + |\g|^2 + \dots,
\ee
and $q$ characterizes the contributions from magnification 
and size bias (\cite{paperI,paperII}).  Specifically, the parameter $q$ is given 
in terms of 
$\beta_f$ and $\beta_r$, the logarithmic slopes of the flux and size 
distributions, as:
\begin{align}
q = 2\beta_f +& \beta_r - 2 \\
\beta_f \equiv -\frac{d\ln \nobs}{d\ln f}\Big\vert_{f=f_{\rm min}},\quad&
\beta_r \equiv -\frac{d\ln \nobs}{d\ln r}\Big\vert_{r=r_{\rm min}}.
\end{align}
Here $f$ denotes flux and $r$ stands for apparent size of the galaxies.  
For definiteness, we assume a value of $q=1.5$, in the middle of the range
estimated by \cite{paperI}.
Consider now the expectation value of $\Aest$ in the presence of lensing bias.
Inserting \refeq{nobs1} into \refeq{Aest1} and taking the expectation value we find
\be
\avg{\Aest} = \int d^2\theta\ \mu^{q/2}(\vth)g(\vth) W(\vth) \\
\ee
where we have assumed the lensing field does not correlated with the source density
field $\d_g$.   We see that the increase in the number of background sources (assuming $q>0$)
leads to a higher signal, as expected.

We turn now to estimating the variance of $\Aest$ in the presence of magnification
bias and source clustering.   We assume that a weak lensing shear peak
occurs when $\Aest$ is evaluated at the center of a lensing halo.  For now, 
we assume that there is no contribution to the lensing
shear and magnification from other matter along the line of sight.  
The clustering of the source population which we neglected in \refsec{est}  
is an additional 
noise contribution to $\Aest$.  Using \refeq{nobs1}, we have
\be
\avg{\d_i\d_j} = \frac{1}{\bar n\:\mu^{q/2} \D\Omega}\delta_{ij} + \xi_{ij}
\ee
where $\xi_{ij} = \xi(|\vth_i-\vth_j|)$ is the source galaxy angular correlation function.  
Here, we neglect higher moments of the angular distribution of source galaxies,
which can become relevant on very small scales.   In order to calculate
$\xi$, we make another approximation: we assume that galaxies follow the 
matter distribution with a linear bias of $\sim 1$, and use the fitting 
formula of \cite{SmithEtal} for the non-linear
matter power spectrum together with the redshift distribution \refeq{dNdz}
(see also appendix of paper~I).  
Note that $\xi$ is the \emph{observed} correlation function, which in principle
is also modified by lensing bias induced by large-scale structure 
\citep{Matsubara2000,HuiGaztanagaLoVerde2007}.  Since this is a small ($< 10$\%) 
correction to a usually subdominant noise contribution, we neglect this
lensing bias here given our rough approximations for $\xi$.  However, we do 
take into account that magnification modifies the source density
through lensing bias, and correspondingly affects the shot noise.  

The mean and variance of source ellipticities remain 
unchanged by lensing bias, so that as before
$\avg{e_i}=g_i$ and $\avg{e_ie_j}=\frac{\sigma_e^2}{2}\delta_{ij}+ g_ig_j$.

Putting everything together and plugging into \refeq{Aest2} we find
\bea
\Var(\Aest) & = & V_{\rm shot} + V_{\rm src}\label{eq:VA}\\
V_{\rm shot} &=& \frac{1}{\bar n} \int d^2\theta\ \mu^{q/2}(\vth) W^2(\vth) \left( \frac{\sigma_e^2}{2}+ g^2(\vth)\right)\  \   \label{eq:Vshot}\\
V_{\rm src} &=& \int d^2\theta \int d^2\theta'\ S\: S'\: \xi(|\vth-\vth'|),
\label{eq:Vsrc}
\eea
where
\be
S(\vth)=\mu^{q/2}(\vth)g(\vth)W(\vth),
\label{eq:S}
\ee
and primed and un-primed variables are evaluated at $\vth'$ and $\vth$ respectively.  We have split the variance of $\Aest$ into a shot-noise contribution
$V_{\rm shot}$, and a contribution from source clustering $V_{\rm src}$.  
Comparing \refeq{VA} with \refeq{Vsh0} we see that lensing bias increases 
the shot noise in $\Aest$ due to the increased source density (but not
as fast as it increases the value of $\Aest$ itself).  In addition, 
the clustering of source galaxies adds to the variance of $\Aest$. 

\begin{figure}[t!]
\centering
\includegraphics[width=0.48\textwidth]{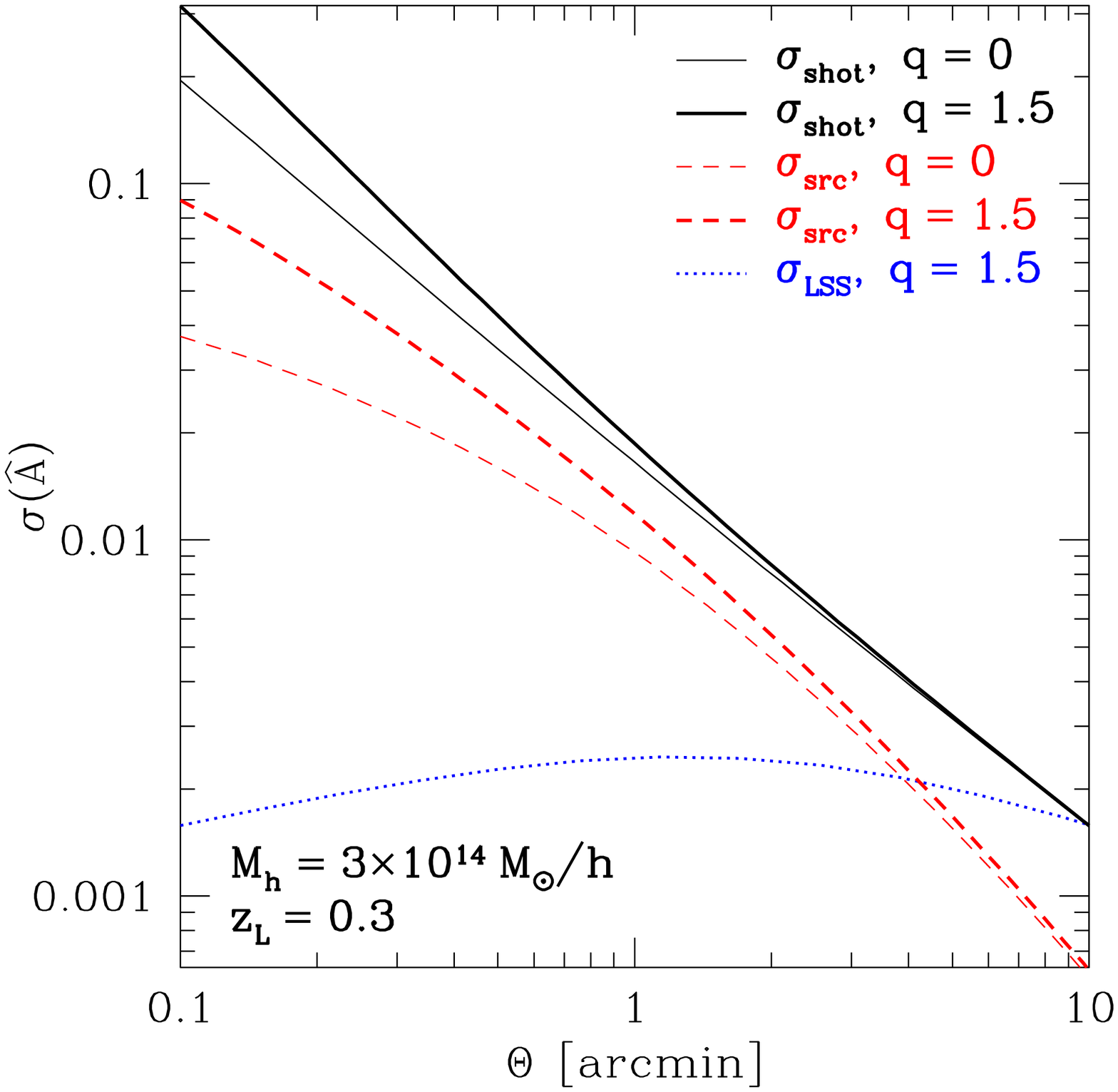}
\caption{Contributions to the variance of smoothed shear estimators as
a function of filter scale $\Theta$ for a Gaussian shear filter:
shot noise $\s_{\rm shot}=\sqrt{V_{\rm shot}}$ [\refeq{Vshot}], source 
clustering noise $\s_{\rm src}=\sqrt{V_{\rm src}}$ [\refeq{Vsrc}], 
and large-scale structure variance $\s_{\rm LSS}$ (\refapp{LSS}).  
$\s_{\rm shot}$ and $\s_{\rm src}$ are shown with (thick lines; $q=1.5$) 
and without (thin lines; $q=0$) lensing bias, while $\s_{\rm LSS}$ is only
shown including the very small lensing bias correction (again for $q=1.5$).  
\label{fig:VA}}
\end{figure}

\reffig{VA} shows the two contributions to the variance of $\Aest$ for a Gaussian
filter [\refeq{WGauss} in \refsec{filter}] as function of the filter scale $\Theta$.  The thick
lines show results including lensing bias ($q=1.5$), while the thin lines
are without lensing bias.  
For the results with lensing bias, we have
assumed that the estimator is centered on an NFW lensing halo 
of mass  $3\times 10^{14}\ \Msunh$ at redshift $z_L=0.3$.  We see that 
source clustering is subdominant compared to shot noise for an average
source density of $\nbar = 20$~arcmin$^{-2}$.  For higher source densities,
source clustering will become important.  We also see that lensing
bias increases both sources of noise, with $V_{\rm src}$ being affected more
strongly.  

Finally, we consider one more source of variance for shear estimators: that
from the lensing field induced by large-scale structure itself.  Generally,
massive dark matter halos or density peaks reside on average in overdense
regions.  This \emph{associated} large-scale structure (LSS) can add to or 
subtract from the lensing 
signal of the halo itself.  Unfortunately, calculating this contribution
properly is only possible using N-body simulations (especially when taking
into account lensing bias).  What we can calculate however is the estimated
variance of $\Aest$ induced by \emph{uncorrelated} large-scale structure,
$\s_{\rm LSS}$, as
detailed in \refapp{LSS} \citep[see also][]{hoekstra01}.
The result is shown as dotted line in \reffig{VA}, again for the Gaussian
shear filter.  Clearly, large scale 
structure noise is subdominant for this filter as long as $\Theta$ is not 
very large;  still, for $\Theta\gtrsim 3$~arcmin, it cannot be neglected.  
Furthermore, the magnitude of $\s_{\rm LSS}$ depends on the type of filter
chosen: for the KS-Gaussian filter, the large-scale structure noise is
much more significant, at the percent level for $\Theta \lesssim 1$~arcmin.  

\subsection{Another Shear Estimator}
\label{sec:Aprime}

An alternative to \refeq{Aest} as choice of smoothed shear estimator uses a 
location-based normalization:
\be
\Best(\vth) = \frac{\sum_i e_i W(\vth_i-\vth)}
{\sum_i W(\vth_i-\vth)} \label{eq:Aprime},
\ee
where both sums run over all galaxies.  
The normalizing denominator removes some of the fluctuations in the
source galaxy density (which can be intrinsic or survey-specific, such as
varying depth of the observations).  We can write \refeq{Aprime} as
\bea
\Best(\vth) &=& \frac{\Aest(\vth)}{\hat N(\vth)},\\
\hat N(\vth) &=& \frac{1}{\nbar} \sum_i n_i W_i \D\Omega\vs
&=& \int d^2\vth' \mu^{q/2}(\vth') [1 + \d(\vth')] W(\vth-\vth'),
\eea
where in the last line we have employed the continuum limit.

It is worth pointing out some differences between $\Aest$
and $\Best$ and their response to systematic effects due to source 
clustering.  For instance, fluctuations in the number density
of galaxies \emph{behind} the lens are canceled out in $\Best$, while they
contribute to $\Aest$.  On the other hand,
if there is an overdensity of galaxies associated with the lens or
in the foreground (present in the sample for example due to uncertainties 
in photometric redshifts), these galaxies systematically reduce the value of $\Best$ 
since they contribute zero shear to the numerator in $\Best$, despite being
included in the denominator.
This
so-called dilution \citep{Bernardeau98,MedezinskiEtal07} does not directly affect the estimator
$\Aest$.  A detailed discussion of the different systematics induced
in both estimators by photometric redshift uncertainties can be found 
in paper~I.

\begin{figure}[t!]
\centering
\includegraphics[width=0.48\textwidth]{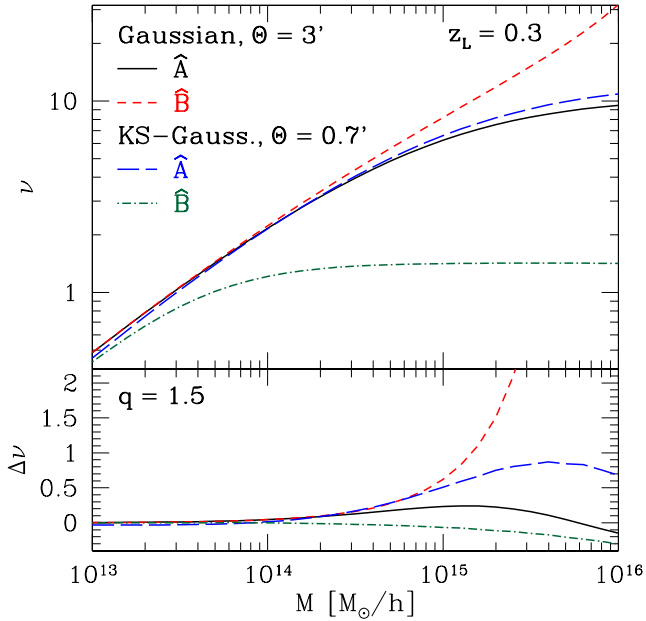}
\caption{\textit{Top panel:} signal-to-noise of estimators $\Aest$ and $\Best$ 
without lensing bias for a lens halo of mass $M$ at $z_L=0.3$, using a
Gaussian shear filter and a KS-Gaussian filter (see \refsec{filter}).  
\textit{Bottom panel:} Change in signal-to-noise 
of $\Aest$ and $\Best$ induced by lensing bias, for the same filters and
lensing halo.  
\label{fig:nu}}
\end{figure}

When calculating the statistics of $\Best$, we face the problem that
$\<\Aest / \hat N\> \neq \<\Aest\>/\<\hat N\>$ (see also paper~I).  
However, we can employ this approximation if the effective number of galaxies
used in the shear estimator, $\sim \nbar \Theta^2$ where $\Theta$ is the filter
scale, is much larger than unity.  
In this case, we have for the expectation value
\be
\<\Best\> = \frac{\<\Aest\>}{\<\hat N\>} = 
\frac{\int d^2\th\:g(\vth) \mu^{q/2}(\vth) W(\vth)}
{\int d^2\th\:\mu^{q/2}(\vth) W(\vth)},
\ee
where we have again assumed that the shear is uncorrelated with source density.  
We see that the effect of lensing bias on $\Best$ is partially canceled by
the denominator, and therefore the expectation value for $\Best$ is only weakly
dependent on lensing bias corrections.  In the absence of lensing bias and systematics, 
$\<\Best\>=\<\Aest\>$.  

When calculating the variance of $\Best$ however, it is
necessary to take into account the covariance between numerator
and denominator.  The full expression for $\Var(\Best)$ is derived
in \refapp{Aprime}, and is given in equation \ref{eq:VAprime2}.
The gist of it is that fluctuations in the
number of source galaxies, due to both shot noise and source clustering, 
are partially canceled out (but see below).  This becomes
especially important for high masses, where the source clustering contribution
to the variance begins to dominate.  Finally, the large-scale structure
variance is the same for both estimators, assuming that it is dominated
by the weak lensing regime (lowest order in $\k$, $\g$).  

The top panel of \reffig{nu} shows the average expected signal-to-noise 
$\nu$, defined as 
$\nu = \<\Aest\>/[\Var(\Aest)]^{1/2}$ for both $\Aest$
and $\Best$, as a function of halo mass.   
We have again assumed an NFW lensing halo, and
we show results both for a Gaussian shear filter of width $\Theta=3'$ 
and a KS-Gaussian filter with $\Theta=0.7'$ (the choices of $\Theta$ will 
be justified in \refsec{filter2}).  Further, we have 
ignored lensing bias for the moment.  For the Gaussian filter, we find that
$\Best$ is superior to $\Aest$ in terms of signal-to-noise for massive halos.  
For lower mass halos, where $g^2 \ll \sigma_e^2$, both estimators are
equivalent.  Similar conclusions hold for the NFW-optimized 
filter.\footnote{In paper~I, we found
that both types of estimators are statistically equivalent.  This is because,
first, source clustering is negligible for the thin annuli considered in the
stacked weak lensing context; and second, for the relevant radial scales,
the stacked analysis is in the regime of $g^2 \ll \sigma_e^2$.}

The KS-Gaussian filter shows a very different behavior: here, $\Best$ has
significantly less signal-to-noise than $\Aest$ at all relevant masses
(in fact, $\nu(\Best)$ never exceeds $\sim 1.4$), 
while $\Aest$ performs very similarly to the corresponding Gaussian
estimator.  The reason is that the covariance between the numerator and 
denominator of $\Best$, which leads to the partial cancelation of fluctuations for the
other shear filters, is strongly suppressed for a KS-Gaussian filter.  
This is a consequence of the inclusion of very large scales in the filter (see \refapp{Aprime}).  
Without significant covariance between numerator and denominator, $\Best$
is just the ratio of two noisy quantities, and not surprisingly has less
signal-to-noise than $\Aest$.  However, as mentioned in \refsec{filter}, this
filter is commonly used on a pixelized shear map, rather than on the
galaxies directly.  Hence, \refeq{Aprime} is actually not used with 
the KS-Gaussian filter in practice.  Nevertheless, this result
illustrates an important point:  
the choice of estimator (e.g., $\Aest$ vs $\Best$) and filter function $W$
has to be done jointly, as the two are interrelated.

\subsection{Impact of Lensing Bias}

We now turn to the impact of lensing bias on $\Aest$ and $\Best$.  
Lensing bias changes both the value (signal) of smoothed shear estimators,
as well as the signal-to-noise.  The latter effect is shown in the 
bottom panel of \reffig{nu}.  Let us discuss the Gaussian filter first.  
We see that for $M\lesssim 3\times 10^{15}\Msunh$,
lensing bias tends to increase the signal-to-noise in both estimators,
an effect which increases with mass.  The effect is in fact greater
for $\Best$ than for $\Aest$, even though the effect on the value of 
$\Best$ is smaller.  This is because lensing bias acts to decrease the
variance of $\Best$ in two ways:  first, the increased source density
reduces shot noise;  second, lensing bias increases the covariance
between numerator and denominator, improving the cancelation of fluctuations
in the galaxy number above that without lensing bias.  
In case of $\Aest$, we see a reversal of the
effect at very high masses: lensing bias in fact \emph{reduces} the 
signal-to-noise of very high peaks for $\Aest$.  This is because for very 
high halo masses (which are necessary to produce this high signal-to-noise), 
source clustering takes
over as the dominant source at this filter scale.  While the source clustering
noise scales similarly with the lensing quantities as the signal itself
($\propto \mu^{q/2} g$), it is weighted by $W^2$ rather than $W$;  this 
weighting favors smaller radii where the lensing magnification is stronger
(see \reffig{kprof} in \refapp{NFW}), thus boosting the noise more than the
signal.  Note however that for such halos the optimal filter scale is 
significantly larger than the assumed 3~arcmin.  On the other hand,
fluctuations due to source clustering are largely canceled out in $\Best$, 
and we do not see this reversal for this estimator.  Instead, the boost 
in signal-to-noise grows strongly towards larger halo masses.  

Again, the KS-Gaussian filter behaves differently: a significant increase
in $\nu(\Aest)$ is seen, while the effect on $\nu(\Best)$ is a small
decrease.  The causes for the differences are, for $\Aest$, that source clustering
is somewhat smaller for the KS-Gaussian filter than for the Gaussian filter,
and for $\Best$, that lensing bias increases the variance of the estimator
slightly faster than it increases the value of $\Best$ itself.  

\begin{figure}[t!]
\centering
\includegraphics[width=0.48\textwidth]{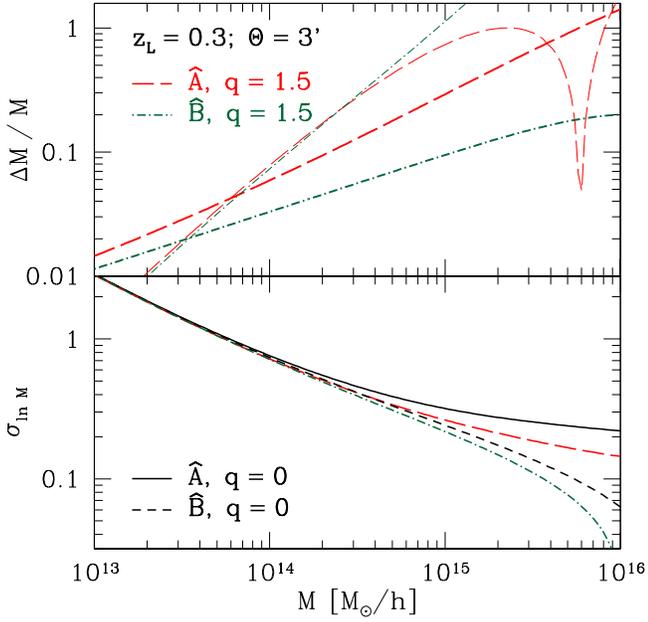}
\caption{Estimated mass bias when neglecting lensing bias 
(top) and relative error (bottom) on the mass 
measurement using filtered shear estimators $\Aest$ and $\Best$.  
In the top panel, the thick lines
show the mass bias expected if the signal itself is used to estimate the
halo mass (thick lines), and if the signal-to-noise is used for the estimate
(thin lines).  In all cases, we have assumed a lens redshift
of $z_L=0.3$, and a Gaussian shear filter with $\Theta=3'$.  
\label{fig:sigmaM}}
\end{figure}

It is also interesting to consider what impact lensing bias has on 
the mass that one would assign to each shear peak.  Assuming one can predict
a relation $\Aest = \Aest(M)$, then from the amplitude of the shear peak one
can estimate its mass.  If, however, one fails to account for the boost in 
signal due to lensing bias, one will systematically overestimate the 
corresponding halo mass.\footnote{Note that in order to predict a mass--shear-peak relation
one needs to assume a lens redshift as well as a profile shape, i.e. halo
concentration.  Here, we assume the true values of $z_L$ and $c$ are known.}  The systematic mass offset is approximately given by 
\be
\frac{\D\Mvir}{\Mvir} = \frac{d\ln M}{d\ln\Aest} \frac{\D\Aest}{\Aest},
\label{eq:DM}
\ee
where the logarithmic slope $d\ln\Aest/d\ln M \approx 0.5-0.7$ depending on 
mass, and correspondingly for $\Best$.  
The top panel of \reffig{sigmaM} (thick lines) shows the relative 
bias $\D M/M$ obtained with a Gaussian filter with $\Theta=3'$.  
Clearly, for the most massive halos biases as large as tens of percent are 
possible.  Note that the bias in $\Best$ is {\it smaller} than that of $\Aest$.  
This is because we are only relying on the amplitude of $\Best$ to estimate a 
cluster's mass, and as we have seen, the amplitude of $\Best$ is less 
affected than that of $\Aest$.  The thin lines in the top panel
of \reffig{sigmaM} show the expected bias if one would instead use
the signal-to-noise of a lensing peak to estimate the halo mass, calculated
using a similar relation to \refeq{DM}, but for $\nu(\Aest), \nu(\Best)$ 
instead $\Aest, \Best$.   Here,
the situation is reversed at high masses:  the mass bias is larger for
$\Best$ than for $\Aest$, reaching order unity at $M\approx 10^{15}\Msunh$.  
Moreover, the bias in $\Aest$ changes
sign at very high masses (for a fixed filter scale).  This reflects the
behavior of the lensing bias effect on $\nu$ seen in \reffig{nu}.

Turning now to the statistical uncertainty in the recovered mass,
we have seen that lensing bias helps increase the signal-to-noise
of a given halo.  Thus, we expect properly accounting for lensing
bias will reduce the statistical error in cluster mass estimates.
The expected error in log-mass can be estimated as
\be
\s_{\ln M} = \frac{d\ln M}{d\ln\Aest} \frac{1}{\nu_A}.
\ee
This is shown in the bottom panel of
\reffig{sigmaM}.  Again, the improvement in mass resolution becomes
relevant at the high-mass end.  We also see how the better statistical
performance of $\Best$ over $\Aest$ for this filter reflects in the smaller mass
uncertainty at the high-mass end.  Note that this should only be seen as rough
estimate; in practice, halo triaxiality and
projection effects will increase the error in the recovered masses significantly.

\begin{figure}[t!]
\centering
\includegraphics[width=0.48\textwidth]{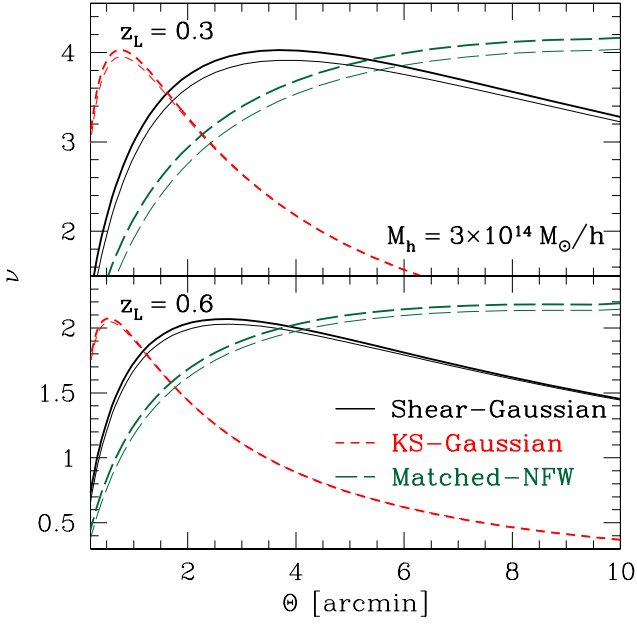}
\caption{Signal-to-noise of the estimator $\Aest$ as function of filter
scale for different filter shapes (truncation scale in case of the matched
NFW filter), for a $3\times 10^{14}\Msunh$ halo at
two different redshifts: $z_L=0.3$ (top panel) and $z_L=0.6$ (bottom panel).  
The thin lines show results without lensing bias, 
while the thick lines include lensing bias ($q=1.5$).  
\label{fig:nu-sigma}}
\end{figure}

\subsection{Choice of Filter Scale}
\label{sec:filter2}

In order to optimize the filter shape, one has to adopt a metric with
which different filtrs can be compared.  
In the following, we will focus on the goal of maximizing the signal-to-noise
for a given lensing halo, which is most directly relevant to shear peak counts.  

In order to test the relative performance of the different filters,
we show the signal-to-noise $\nu$ for a $3\times 10^{14}\Msunh$ lensing
halo at fixed lens redshifts of $z_L=0.3,\; 0.6$ in \reffig{nu-sigma}, as a 
function of the filter scale $\Theta$.  In case of the matched-NFW filter,
$\Theta = \theta_{\rm max}$ is the truncation scale of the filter.  
We use $\Aest$ in all cases.  The optimal signal-to-noise
is determined by a balance between the signal $\Aest$, which decreases
with increasing filter size, and the noise which also decreases.  
We see that all three filters
peak at roughly the same signal-to-noise for $\Aest$, though the filter scale at
which this peak is reached is very different.  For the Gaussian filter,
the optimal case is around $\Theta \approx 3'$, which we have thus adopted
as default value of the filter scale.  For the KS-Gaussian, the optimal
value is $\Theta \approx 0.7'$.  Note that these values depend on the mass
and redshift of the lens.  

The relative importance of the three noise contributions, shot noise,
source clustering, and LSS noise, depends on the filter shape.  In
case of the KS-Gaussian filter, the LSS noise is much more significant,
a factor of $\sim 10$ higher at the optimal filter scale than for the
other two filters.  Again, this is a consequence of the inclusion of
very large scales in this filter.  This will also be of relevance
to the contribution of correlated large-scale structure to smoothed
shear estimators.

\reffig{nu-sigma} also shows the effect of lensing bias 
(thick lines vs thin lines):  
for this halo, lensing bias boosts the peak signal-to-noise by 
$\sim 5$\% for the Gaussian and NFW-optimized filters.  In case of the
KS-Gaussian, the effect is slightly smaller, $\sim 3$\%, due to
the preference for large scales in that filter.  Note that the optimal 
filter scale is moved to
slightly smaller values by lensing bias.  This is because the signal in
$\Aest$ increases more steeply with decreasing filter size when including
lensing bias.  

Before deciding on an optimal filter, however, it is necessary to take
into account the effect of associated and coincidental large-scale structure
along the line of sight to the lens.  Clearly, the likelihood of a chance
superposition with unrelated matter concentrations with the filter scale 
\citep{Hoekstra02,MaturiEtal05}.  This might make a true optimal filter
narrower than suggested by \reffig{nu-sigma}.  


\section{The Peak Function}
\label{sec:res}

We now turn to investigating the impact of lensing bias and filter choice
on the abundance of detected shear peaks.  
To do so, we assume each shear peak corresponds to an NFW halo
and that the peak position is the halo center.  While in practice
one expects some fraction of weak lensing peaks to arise due
to chance superpositions of multiple halos, the results we obtain concerning
how lensing bias impacts weak lensing peak finding should be
indicative of the whole population.  

\begin{figure}[t!]
\centering
\includegraphics[width=0.48\textwidth]{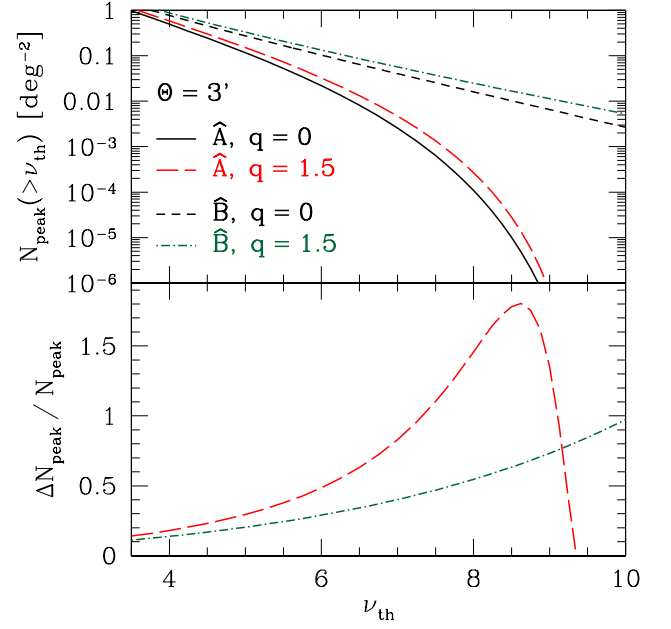}
\caption{
\textit{Upper panel:} Average number peak of peaks $\Np(>\nuth)$ 
(per deg$^2$)
above the signal-to-noise threshold $\nuth$, with and without lensing
bias for the two estimators $\Aest$, $\Best$.  We set $\Theta=3$~arcmin.  
\textit{Lower panel:} Relative lensing bias effect on the peak counts,
$\D \Np(>\nuth)/\Np(\nuth)$.
\label{fig:Npeak}}
\end{figure}

We estimate the average number of lensing peaks within a solid
angle $\Omega_s$ above a given
signal-to-noise threshold $\nuth$ as follows:
\begin{align}
\Np(>\nuth) =\:& \Omega_s\int_0^{\infty} \frac{c\:dz_L}{H(z_L)}\chi^2(z_L)\label{eq:Npeak}
\\
& \times\int_{\ln M_{\rm min}(z_L,\nuth)}^{\infty} d\ln M\:\frac{dn}{d\ln M}(z_L,M),\quad\nonumber
\end{align}
where $c$ is the speed of light, $H(z)$ is the Hubble expansion rate,
$\chi(z)$ is the comoving distance to redshift $z$, $dn/d\ln M$ is the 
halo mass function, and $\Mmin$ is defined via
\be
\nu(\Mmin, z_L) = \nuth.
\label{eq:Mmin}
\ee
Hence, lensing bias enters by
lowering the effective $z$-dependent mass threshold of the survey, so that
the number of peaks is increased.  We use the fitting function
of \cite{TinkerEtal} to calculate $dn/d\ln M$ as function of mass and redshift
from the linear matter power spectrum (recall that $M=M_{200m}$ throughout).

\reffig{Npeak} shows the peak count statistics $\Np(>\nuth)$ (peaks per deg$^2$)
with and without lensing bias using a Gaussian filter with
fixed filter scale $\Theta = 3$~arcmin.  We also show the
relative lensing bias effect $\D \Np/\Np$.  Evidently, estimator $\Best$
is more efficient at finding peaks for this filter, with or without lensing 
bias: the number of high signal-to-noise peaks ($\nu \gtrsim 6$) is higher
than that found by $\Aest$ by orders of magnitude.  
Further, lensing bias can boost the weak lensing peak counts significantly
for both $\Aest$ and $\Best$, by more than a factor of 2 for high peaks.  
At high thresholds $\nuth \gtrsim 8.5$, we see a turnover, which 
reflects the trend seen in $\Delta\nu$ (\reffig{nu}).  

\begin{figure}[t!]
\centering
\includegraphics[width=0.48\textwidth]{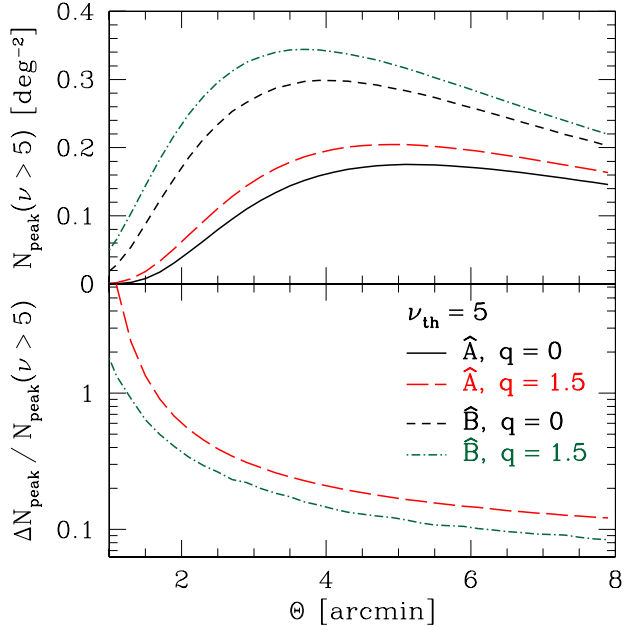}
\caption{\textit{Upper panel:} Average number peak of peaks $\Np(\nu>5)$ 
(per deg$^2$), with and without lensing bias for the two estimators 
$\Aest$, $\Best$ as a function of the filter scale $\Theta$.
Other survey specifications as in \reffig{VA}.  \textit{Lower panel:}
Relative lensing bias effect on the peak counts,
$\D \Np(\nu>5)/\Np(\nu>5)$.
\label{fig:Npeak-sigma}}
\end{figure}

Note that for very high peaks, i.e. massive halos, a filter scale of 
3~arcmin is not optimal.  Hence, it also interesting to consider 
the number of peaks above a fixed threshold
as function of the filter width (\reffig{Npeak-sigma}).  The different
number of peaks here reflects the different mass thresholds for different filter
scales at a fixed signal-to-noise threshold.  Again, the location-normalized
estimator $\Best$ yields more peaks than $\Aest$,  and lensing bias
increases the number of $5\sigma$ shear peaks found with either estimator
by $\sim$30\%.   Note that since lensing
bias pushes the optimal filter scale to smaller values, the number of peaks 
increases faster with scale when incorporating
the impact of lensing bias.  Not surprisingly, we find that choosing smaller filters
can increase the relative importance of lensing bias significantly
(see the lower panel of \reffig{Npeak-sigma}).

Finally, we note that the result of the comparison $\Aest$ vs $\Best$ 
reverses for the KS-Gaussian filter:  $\Aest$ performs far better for this
filter, as expected after \refsec{Aprime}.  

\section{Summary and Discussion}
\label{sec:disc}

We have compared the expected signal-to-noise of the two most common
types of filtered shear estimators for different filter functions,
and studied how lensing bias impacts these estimators.  
In our signal-to-noise
considerations, we also include the effects of the intrinsic clustering
of source galaxies, and the variance of the estimators due to uncorrelated
large-scale structure.  The former noise contribution becomes important
for massive lensing halos, while the latter's importance only depends on the
filter shape and scale considered.  

We find that estimator and filter function need to be chosen jointly
and cannot be regarded as independent.  For example, for the Gaussian
shear filter, the location-normalized estimator $(\Best)$ is statistically
superior to the globally normalized estimator $(\Aest)$ for high peaks.  
This is because fluctuations in the number of galaxies are
canceled out to first order in $\Best$.  For the KS-Gaussian filter
on the other hand, the situation is reversed:  $\Aest$ performs far
better than $\Best$.  According to our (certainly not exhaustive) results, 
a location-normalized estimator with a Gaussian-type shear filter 
appears to perform best statistically.  The question of optimal
filter+estimator combination certainly deserves more attention, as 
the abundance of high peaks can be suppressed by orders of magnitude
for suboptimal choices (\reffig{Npeak}).  

Another finding, of equal importance to their statistical properties, 
is that the two types of estimators respond
very differently to uncertainties in the photometric
redshifts (see the discussion in paper I).  Specifically,
the globally normalized estimator $\Aest$ does not suffer from the
so-called dilution effect affecting $\Best$, and should be much less 
sensitive to contamination of the source sample by galaxies associated
with the lens.  

Turning to lensing bias, we find that it affects both of the estimators 
we considered, and for all filter functions.  While the signal-to-noise 
of either estimators can be boosted significantly (up to $\Delta\nu\sim 1-2$), 
the value of the estimator $\Aest$ is generally affected more 
strongly than that of $\Best$.  Indeed, if one were to estimate halo masses 
based solely on their smoothed shear signal, halo mass over-estimates as large 
as tens of percent are possible.   Not surprisingly, the increase in 
signal-to-noise especially of high peaks also results in 
comparable boosts to the abundance of observed peaks.  The magnitude of the 
effect depends strongly
on the filter scale as well:  smaller filter scales lead to a much
larger boost in signal due to lensing bias, pushing the optimal filter
scale towards smaller values.  

While we have not considered the impact of halo triaxiality and correlated structures in our work,
it is straightforward to understand how these can affect our conclusions.  Specifically, both
of these sources of noise tend to increase the weak lensing shear signal, and therefore
will increase the relative importance of lensing bias.
Moreover, if one wishes to minimize line-of-sight projections of multiple halos, we expect doing so
will require smaller filters, resulting in a further increase of the importance of lensing bias.

In light of these results, it is clear that a proper modeling of lensing bias 
as well as source clustering 
will be a necessary component of cosmological interpretations of 
the shear peak function.  Furthermore, these effects should also be taken
into account when designing an optimal estimator+filter combination 
for shear peak finding.  
Fortunately, incorporating these two effects is fairly
straightforward, both 
for analytic calculations and numerical studies with N-body simulations.  
Lensing bias, by increasing the signal-to-noise of the shear signal, has
the potential to significantly boost the 
statistical power of shear selected samples of objects.  Thus, properly 
accounting for lensing bias should allow us to maximize the cosmological 
potential of shear peak statistics.


\acknowledgments
We would like to thank Richard~Ellis, Richard~Massey, James~Taylor, and Scott Dodelson for 
helpful discussions.  

F.~S. is supported by the Gordon and Betty Moore Foundation at Caltech.  E.R.  is funded by
NASA through the Einstein Fellowship program, grant PF9-00068.

\begin{appendix}

\section{Lensing by an NFW halo}
\label{app:NFW}

In this appendix we review our lens model used for the numerical results.  
The Navarro-Frenk-White density profile is given by \citep{NFW}:
\be
\rho(r) = \frac{\rho_s}{r/r_s (1+r/r_s)^2},\quad \rho_s = \frac{M}{4\pi\:r_s^3 [\ln (1+c) - c/(1+c) ]},
\ee
where $r_s$ is the scale radius, and $c=R_{200}/r_s$ is taken from the fit 
of \cite{BullockEtal}.  Note that 
this halo profile is not truncated at $R_{200}$.  Actual halo profiles lie 
somewhere in between the extrapolated NFW profile adopted here and
a truncated profile.  The differences however appear mainly around 
$R_{200}$, which is typically larger than the filter size we will consider.  
Hence, the details of the outer halo profile do not change the results 
significantly.  

Further, we make the small angle approximation.  
The lensing quantities $\k$ and $\gamma$ for an NFW
halo have been derived in the literature (\cite{WrightBrainerd,MaturiEtal05}).  
For reference, the convergence at the scale radius is given by
\be
\k_s = \frac{\rho_s r_s}{\Sigma_{\rm cr}} = \frac{3}{2}\frac{\rho_s r_s H_0^2}{\rho_{\rm cr,0}c^2} (1+z_L) W_L(z_L).
\ee
Here, $z_L$ is the halo redshift, 
$\rho_{\rm cr,0}$ is the critical density today, $c$ is the speed of light,
and the lensing weight function is given by
\be
W_L(z_L) = \frac{c}{H(z_s)}\int_{z_L}^{\infty} dz_s \frac{\chi_L}{\chi(z_s)}
[\chi(z_s)-\chi_L] \frac{dN}{dz_s},
\ee
where $\chi_L=\chi(z_L)$,
and $dN/dz$ is the normalized redshift distribution of the source galaxies,
\refeq{dNdz}, determined from a fit to the expected redshift distribution of 
galaxies for DES \citep{Annis}.

\begin{figure}[t!]
\centering
\includegraphics[width=0.48\textwidth]{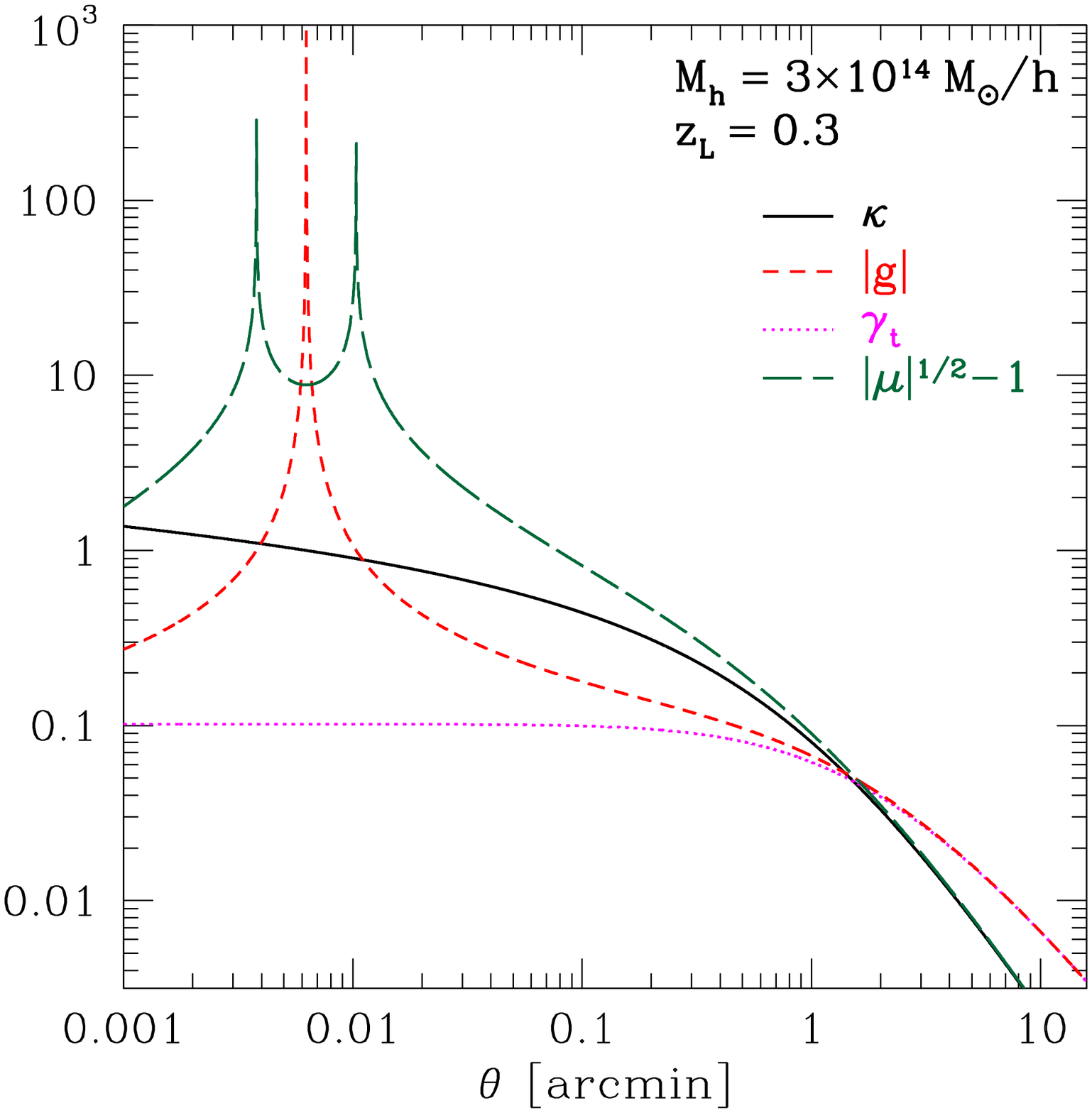}
\includegraphics[width=0.48\textwidth]{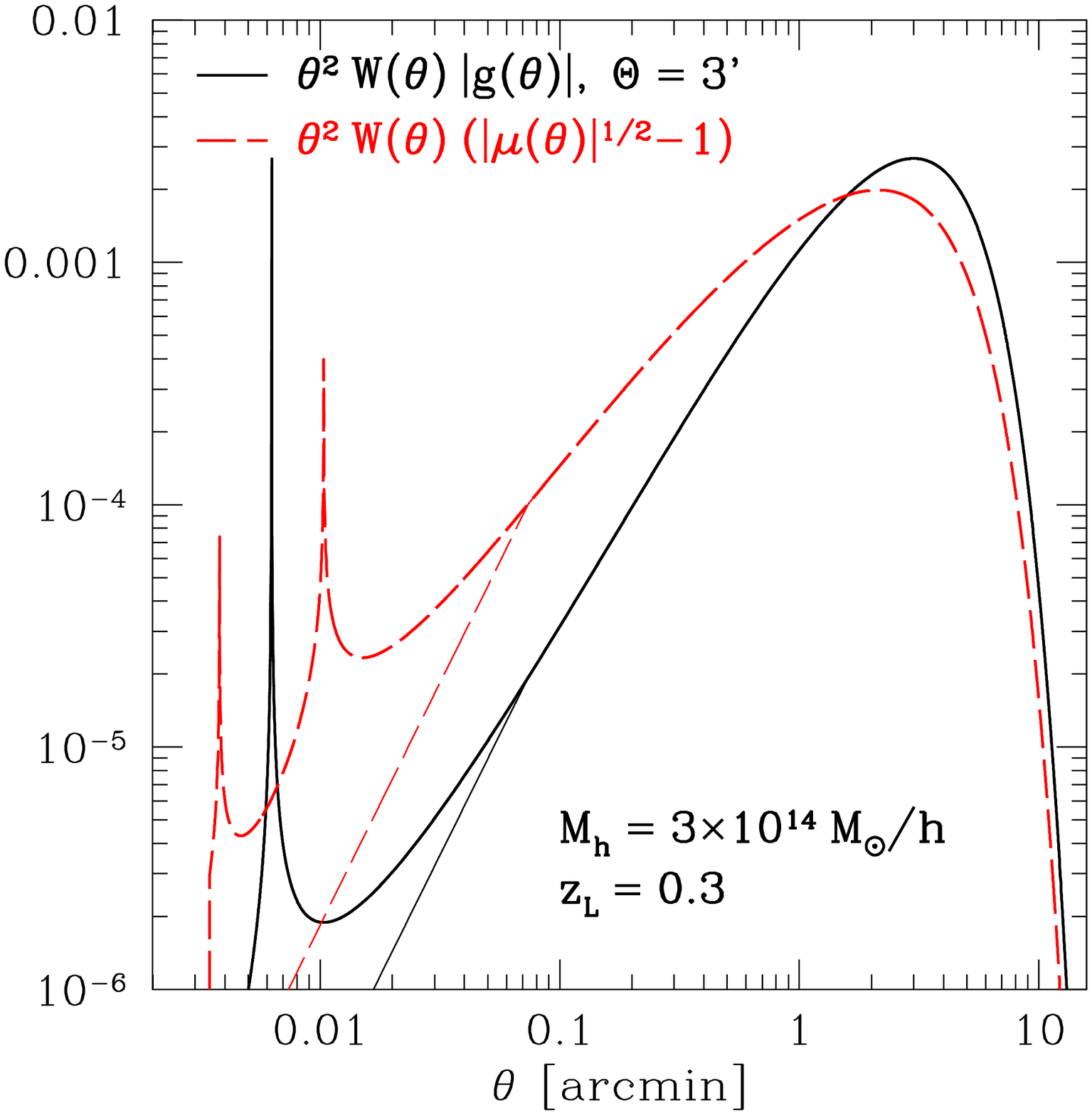}
\caption{\textit{Left panel:} Profile of convergence $\k$, tangential shear $\g$, 
tangential reduced shear $g$ and scaled magnification $\sqrt{\mu}-1$ around a halo of mass
$3\times 10^{14}\Msunh$ located at $z_L=0.3$.  
\textit{Right panel:} Reduced shear and magnification weighted by area
and the Gaussian filter $W$, for the same parameters as in the left panel. 
This shows which scales contribute to the signal
in \refeq{Aest} and \refeq{Aprime}.  The thin lines show the results capped
at $\kappa = 0.5$ (see text).
\label{fig:kprof}}
\end{figure}

\reffig{kprof} (left panel) shows the profile of $\k$, $\g$, $g$ as well as $\mu$ for a typical
detectable lensing halo at $z_L=0.3$.  
While $\k$, $\g$ are well-behaved, both $g$ and $\mu$
show the strong-lensing caustics, as expected. For this halo and source/lens redshifts,
the strong-lensing regime defined by $g(\theta) \sim 1$ is relevant 
for $\th \lesssim 0.05'$. This regime cannot be treated by weak lensing analysis techniques,
so it should be removed. Fortunately, the bulk of the signal-to-noise
for weak lensing is at much larger radii.   
The right panel of \reffig{kprof} shows the relevant lensing quantities weighted
by the Gaussian filter \refeq{WGauss} for $\Theta=3$~arcmin. Clearly, the bulk of the
signal comes from $\theta\sim0.3 - 3$~arcmin, well outside the strong lensing regime
for this halo. 
Although not of practical relevance, very massive halos do appear in the calculations
where the strong lensing regime extends to beyond 0.2~arcmin. For those cases,
we cap the convergence at $\k = 0.5$. This is merely done to lead
to convergence of the calculation, and only occurs for such high halo masses 
that it does not impact the results presented here.

\section{Large-scale structure variance}
\label{app:LSS}

Here we re-derive the variance of $\Aest$ due to large scale structure in
the survey \citep[see][]{hoekstra01}.  Since the bulk of the survey area will have small convergence,
we can work in the weak-lensing limit, in which case $\Aest$ and $\Best$
are equivalent.  Furthermore, we only calculate the leading contribution
from the power spectrum of the lensing field, neglecting higher order moments
which in principle can become relevant on very small scales.  
The computation is most conveniently 
done in Fourier space.  For this, we first rewrite \refeq{Aestc} in the
weak lensing limit ($g=\g$) as
as an expression for the filtered convergence $\k$ 
\citep[e.g.,][]{BartelmannSchneider2001}:
\bea
\Aest(\vth) &=& \int d^2\th' \k(\vth') W^\k(|\vth-\vth'|)\label{eq:Akappa}\\
W^\k(\th) &=& 2 \int_\th^\infty \frac{d\th'}{\th'} W(\th') - W(\th).
\eea
Note the non-local relationship between the shear filter and the
corresponding convergence filter.  
Since $\k$ is a scalar quantity, \refeq{Akappa} can be straightforwardly
written in Fourier space:
\be
\tAest(\vl) = \tilde\k(\vl)\tilde W^\k(\l),
\label{eq:tAest}
\ee
where tilded quantities stand for Fourier transforms:
\be
\tilde X(\vl) = \int d^2\th \:e^{i\vl\cdot\vth} X(\vth).
\ee
Then, using the definition of the angular power spectrum of the convergence:
\be
\<\k(\vl)\k(-\vl')\> = (2\pi)^2 \d_D(\vl-\vl') C^\k(\l),
\ee
we can write the first-order variance of $\Aest$ as:
\be
\s^2_{\rm LSS} = \<\Aest^2\> = \int \frac{d^2\l}{(2\pi)^2} |\tilde W^\k(\l)|^2 \:C^\k(\l).
\ee
This quantity is shown as dotted line in \reffig{VA}, for
a Gaussian shear filter \refeq{WGauss} (note that $W^\k$ for this filter
is not a Gaussian).  
It is also straightforward to calculate the leading lensing bias correction
to $\s_{\rm LSS}$, following the second order
correction to $C^\k(\vl)$ presented in \cite{paperII}.  While we
include this correction, the relative change of $\s_{\rm LSS}$ only amounts 
to a few percent for filter scales $\Theta\gtrsim 1'$  (see also \cite{White05}).  

\section{Variance of the location-normalized estimator}
\label{app:Aprime}

In this section, we derive the variance of $\Best$ due to shot noise
and source clustering, taking
into account the covariance between numerator and denominator.  We
expand $\Best = \Aest/\hat N$ around its expectation
value, $\<\Best\> = \<\Aest\>/\<\hat N\>$, assuming that fluctuations are
much smaller than unity (justified if $\bar n\,\Theta^2 \gg 1$).  
In close analogy to the derivation in the appendix of paper~I, this yields
\be
\frac{\Var(\Best)}{\<\Best\>^2} = \frac{\Var(\Aest)}{\<\Aest\>^2} 
+ 3\frac{\Var(\hat N)}{\<\hat N\>^2} - 4\frac{{\rm Cov}(\Aest,\hat N)}{\<\Aest\> \<\hat N\>},\label{eq:VAprime}
\ee
where
\bea
{\rm Cov}(\Aest,\hat N) \equiv \<\Aest\hat N\> - \<\Aest\>\<\hat N\> 
&=& \frac{1}{\bar n} \int d^2\th\: W^2(\th)\mu^{q/2}(\th) g(\th)
+ \int d^2\th\int d^2\th'\: W \mu^{q/2} W' \mu'^{q/2} g' \xi(|\vth'-\vth|)\\
\Var(\hat N) &=& \frac{1}{\nbar}\int d^2\th\:W^2(\th) \mu^{q/2}(\th) 
+ \int d^2\th\int d^2\th'\: W \mu^{q/2} W' \mu'^{q/2} \xi(|\vth-\vth'|).\quad
\eea
Here, we have used the shorthand notation introduced after \refeq{S}.  
In each of these equations, the first term denotes the shot noise
contribution, while the second is the contribution from source
clustering.  We see from \refeq{VAprime} that there is a cancelation
of noise terms (both shot noise and source clustering) due to the positive 
covariance between $\Aest$ and $\hat N$.  
The cancelation is not perfect, since the various terms involve different
integrals over shear and magnification.  However, 
for relatively low-mass halos for which $g^2 \ll \sigma_e^2$, the shot noise
term of $\Var(\Aest)$ [\refeq{Vshot}] dominates in \refeq{VAprime}.  In this 
limit, we recover $\Var(\Best) = \Var(\Aest)$ (see \reffig{nu}).  

If we neglect lensing bias and source clustering in both $\Best$
and $\Aest$, so that $\<\Aest\> = \<\Best\>$, \refeq{VAprime} simplifies to
\be
\Var(\Best) = \Var(\Aest) + \frac{\<\Aest\>^2}{\bar n\:\A}\left (
3 - 4 \frac{\A \int d^2\th\:W^2 g}{\int d^2\th\: W\: g}\right ),
\label{eq:VAprime2}
\ee
where we have defined $\A = 1/\int d^2\th\: W^2$ as the effective filter 
area.  If the shear $g$ was constant, then the last term in \refeq{VAprime2}
would evaluate to 4, canceling the $g^2$ term in $\Var(\Aest)$ [\refeq{Vshot}]
and leading to $\Var(\Best) = \sigma_e^2/(2\bar n\, \A)$.  
\refeq{VAprime2} makes it clear that the last term determines whether $\Best$ 
performs better or worse than $\Aest$:  if $\A \int W^2 g / \int W\, g$
is order unity, $\Var(\Best) < \Var(\Aest)$.  If it is much less than 1,
$\Var(\Best)$ can be significantly larger than $\Var(\Aest)$.  The latter
is the case for the KS-Gaussian filter, where  $\A \int W^2 g / \int W\, g
\approx 0.2-0.3$, due to the preferential weighting of very large scales.  
Similar reasoning applies to the source clustering contribution.  
This explains why $\Best$ performs worse than $\Aest$ for the KS-Gaussian
filter.  Note that this effect will be mitigated when employing this
filter on a pixelized shear map:  since the sum in \refeq{Aprime} now only 
runs over galaxies within a small pixel (with $W=1$) instead of running
over all galaxies, there will be significant covariance between numerator 
and denominator, reducing the noise in the pixelized shear.  

\end{appendix}

\bibliography{sizebias}

\end{document}